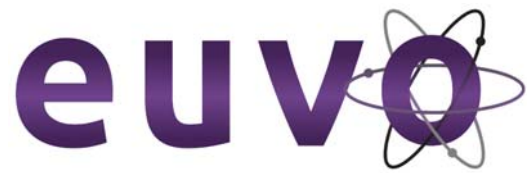

# EUROPEAN ULTRAVIOLET-VISIBLE OBSERVATORY

*"Building galaxies, stars, planets and the ingredients for life between the stars"*

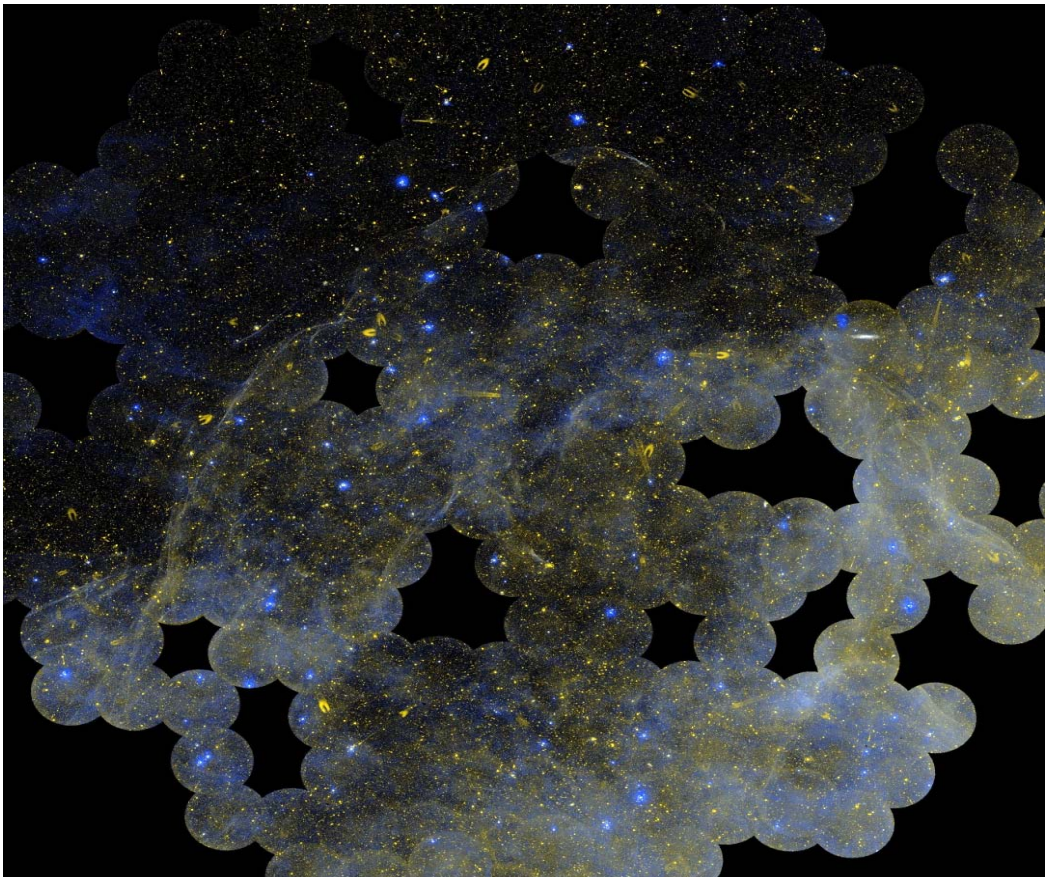

*The UV sky from GALEX All Sky Survey*

Spokesperson: Ana Inés Gómez de Castro
Contact details: AEGORA-Facultad de CC Matemáticas
          Universidad Complutense de Madrid
          Plaza de Ciencias 3, 28040 Madrid, Spain

          email:aig@ucm.es
          Phone: +34 91 3944058
          Mobile: +34 659783338

LIST OF AUTHORS AND CONTRIBUTORS

Thierry Appourchaux – IAS France
Martin Barstow – University of Leicester, United Kingdom
Mathieu Barthelemy - IPAG, France
Fréderic Baudin – IAS, France
Stefano Benetti – OAPD- INAF, Italy
Pere Blay -  Universidad de Valencia,  Spain
Noah Brosch - Tel Aviv University, Israel
Enma Bunce - University of Leicester, United Kingdom
Domitilla de Martino – OAC-INAF, Italy
Ana Ines Gomez de Castro – Universidad Complutense de Madrid, Spain
Jean-Michel Deharveng – Observatoire Astronomique Marseille-Provence, France
Roger Ferlet - Institute d'Astrophysique de Paris, France
Miriam García – IAC, Spain
Boris Gaensicke -University of Warwick, United Kingdom
Cecile Gry - Observatoire Astronomique Marseille-Provence, France
Lynne Hillenbrand – Caltech, USA
Eric Josselin - University of Montpellier, France
Carolina Kehrig – Instituto de Astrofísica de Andalucía, Spain
Laurent Lamy - LESIA, France
Jon Lapington – University of Leicester, United Kingdom
Alain Lecavelier des Etangs – Institute d'Astrophysique de Paris, France
Frank LePetit – Observatoire Paris-Meudom, France
Javier Lopez Santiago - Universidad Complutense de Madrid, Spain
Bruno Milliard- Observatoire Astronomique Marseille-Provence, France
Richard Monier -  Université de Nice, France
Giampiero Naletto – University of Padova, Italy
Yael Nazé - Liège University, Belgium
Coralie Neiner - LESIA, France
Jonathan Nichols – University of Leicester, United Kingdom
Marina Orio – OAPD-INAF, Italy
Isabella Pagano – OACT-INAF, Italy
Céline Peroux – Observatoire Astronomique Marseille-Provence, France
Gregor Rauw – University of Liège, Belgium
Steven Shore – University of Pisa, Italy
Marco Spaans -  Kaptein Astronomical Institute, The Netherlands
Gagik Tovmassian - Instit. Astr. Sede Ensenada-UNAM,  Mexico
José Vilchez – Instituto de Astrofísica de Andalucía,  Spain
Kevin France -  University of Colorado,  USA
Asif ud-Doula - Penn State University, USA

LIST OF SUPPORTERS (see www.nuva.eu/whitepaper/supporters.php)

***Executive summary:***
The growth of luminous structures and the building blocks of life in the Universe began as primordial gas was processed in stars and mixed at galactic scales. The mechanisms responsible for this development are not well-understood and have changed over the intervening 13 billion years. To follow the evolution of matter over cosmic time, it is necessary to study the strongest (resonance) transitions of the most abundant species in the Universe. Most of them are in the ultraviolet (UV; 950 Å – 3000 Å) spectral range that is unobservable from the ground. A versatile space observatory with UV sensitivity a factor of 50-100 greater than existing facilities will revolutionize our understanding of the Universe.

Habitable planets grow in protostellar discs under ultraviolet irradiation, a by-product of the star-disk interaction that drives the physical and chemical evolution of discs and young planetary systems. The electronic transitions of the most abundant molecules are pumped by this UV field, providing unique diagnostics of the planet-forming environment that cannot be accessed from the ground. Earth's atmosphere is in constant interaction with the interplanetary medium and the solar UV radiation field. A 50-100 times improvement in sensitivity would enable the observation of the key atmospheric ingredients of Earth-like exoplanets (carbon, oxygen, ozone), provide crucial input for models of biologically active worlds outside the solar system, and provide the phenomenological baseline to understand the Earth atmosphere in context.

In this white paper, we outline the key science that such a facility would make possible and outline the instrumentation to be implemented.

1. INTRODUCTION

Any future Cosmic Vision requires probing the conditions for the emergence of life in the Universe. In this white paper we present a path to own this future.

For organic molecules to exist nucleo-synthesis needs to have proceeded. Studies of the metal abundance variation up to redshift z=5 demonstrate that the metallicity increases steadily with the age of the Universe. However, the metal enrichment of the Universe was clearly neither uniform nor homogeneous; metal-poor clouds have been detected and chemically processed material has been found in the voids of the Cosmic Web. The star formation rate is observed to decrease from z=1 to the present. Important clues on the metal enrichment spreading on the Universe depend on inter-galactic transport processes; these are poorly studied because of the lack of high sensitivity spectral-imaging capabilities for detecting the warm/hot plasma emission from galactic halos. Current information derives from ultraviolet (UV) absorption-line spectroscopy of the presence of strong background sources. Most of the intergalactic emission is expected to come from circumgalactic filaments and chimneys that radiate strongly in the UV range. To study these structures a high sensitivity spectral-imaging capability is required with spatial resolution at least ten times better than those provided by the GALEX mission.

Metallicity is relevant for life generation as we know it not only at the DNA level but also at much earlier phases. Silicates and carbonates are the key building blocks of dust grains and planetesimals in protostellar/protoplanetary discs. The far UV radiation is a major contributor to disc evolution. It drives the photo-evaporation of the gas disc setting the final architecture of the giant planets in the system and beginning the epoch of rocky planet formation. Unfortunately, little is known about the FUV radiation from solar-system precursors. The measurements carried out from X-ray to softer UV bands point out that the FUV flux varied significantly during the pre-main sequence (PMS) evolution. Protostellar discs are shielded from the energetic stellar radiation during the early phases (<1Myr), but as they evolve into young planetary discs, the FUV and extreme UV (EUV) radiation from the very active young Suns, irradiates them heavily. Strong stellar winds are expected to interact with the left over particles and produce diffuse Helium and Hydrogen emission that pervades the entire young systems during the planets early evolution and planetary atmosphere formation. Around the Sun, within a modest radius of 500 pc, there are thousands of young solar-like stars of all masses and in all the phases of the PMS evolution. The observation of these sources with high sensitivity mid resolution spectroscopy would provide a unique perspective on atmospheres, magnetospheres and coronal evolution, as well as on their impact on planetary formation and evolution.

Solar system planetary research is fundamental for understanding atmospheres as global systems, including the Earth. For example, the link between upper and lower atmospheres is poorly known for the Earth case despite the fact it can have implications on global warming. Studying the upper atmosphere of not only Solar System planets but also of exoplanets can help the understanding of the mechanism operating at Earth[1] [83].

The stellar or solar FUV-EUV fluxes are the main energy input at high atmospheric altitudes. Many atmospheric atoms, ions and molecules have strong electronic transitions in the UV-visible domain. This wavelength range gives access to fundamental constituents of the atmospheres. In particular, bio-markers like Ozone ($O_3$) and molecular oxygen ($O_2$) have very strong absorption transitions in the ultraviolet protecting complex molecules especially DNA from dissociation or ionization. Absorption of $O_3$ through the Hartley bands occurs between 2000-3000 Å and $O_2$ has strong absorptions in the range 1500-2000 Å. Atomic oxygen presents a resonance multiplet at 1304 Å and the famous auroral green and red lines in the visible. For other planetary cases or paleo-Earth, CO has strong absorption bands below 1800 Å and forbidden emission bands from 2000 Å to 3000 Å. Lyman-α (Ly $\alpha$) is the strongest emission line from Earth's atmosphere and is an invaluable tracer for studying the giant planets and hot Jupiters. Hydrocarbons, e.g. methane, are strong FUV absorbers and are sensitive links to the radiative balance in a planetary atmosphere. Observations in the UV and visible wavelength are therefore powerful diagnostics of the structural, thermal, and dynamical properties of planets, be they Solar System or extrasolar.

The UV is an essential spectral interval for all fields in astrophysical research; imaging and spectral coverage at UV wavelengths provides access to diagnostic indicators for diffuse plasmas in space, from planetary atmospheres to elusive gas in the intergalactic medium (IGM). Linking visible and UV spectral features covers the widest possible range of species and vast range of temperatures that cover most astrophysical processes. Moreover, UV observations are essential for studying processes outside of strict thermal equilibrium that produce conditions favorable to complex chemistry, the production pathway for large molecules that absorb and shield planetary surfaces from the harsh space conditions. But UV radiation itself is also a powerful astrochemical and photoionizing agent. Moreover, UV-visible instrumentation provides the best possible spatial resolution for normal incidence optics, since resolution is inversely proportional to the radiation wavelength.

The European community will continue to profit from the refurbished Hubble Space Telescope (HST), a 23 years old space telescope, until the mission end. Following that, the only major telescope world-wide that will provide access to the UV range is the Russian-led WSO-UV (170 cm) space telescope, which includes only minor European participation from Spain. Among the astronomical research lines given the maximum priority in Europe, three require access to the UV range: (1) Planets and Life, (2) The Solar System and (3) The Universe. This even extends to the investigation of the fundamental physical laws, e.g the UV-visible range is without rival for measurement of the variation of the fine structure constant during most of the age of the universe. A versatile UV-visible facility that improves the sensitivity of the dying ultraviolet (UV) facilities by a factor of 50-100 will produce a revolution in our understanding of the Universe. In this White Paper, we outline the key science to be enabled by expanding the astronomical community's access to UV-visible observations in the coming decades.

2. SCIENCE PROGRAM FOR A LARGE ULTRAVIOLET OBSERVATORY

2.1 TRANSPORT PROCESSES IN THE INTERGALACTIC MEDIUM OVER 80% OF THE UNIVERSE LIFETIME

Our Universe is filled by a cosmic web of dark matter (DM) in the form of filaments and sheets. The recent results from the Planck mission [1] show that the DM is one of the two major components of the Universe, forming about 26.8% of its content. The baryonic matter is less than 5%, with the visible baryons being only one-quarter of this amount, ~1% of the total content of our Universe. One important question deriving from this is, therefore, where are most of the baryons.

---

1  Crip, 2013 – www.lpi.usra.edu/vexag/meetings/ComparativeClimatology/presentations/

The DM large-scale mesh was created by primordial density fluctuations and is detected in the present day galaxy distribution, the large-scale structure (LSS). The pattern of density fluctuations was imprinted and can be observed in the cosmic microwave background. At least some baryonic matter follows the DM web components and, in places where the gravitational potential is strong, this baryonic matter collapses to form visible stars and galaxies. Matter not yet incorporated in galaxies is "intergalactic matter" (IGM), with more than 50% of the baryons predicted to be contained at low redshift in a warm-hot phase (WHIM), shock-heated at about $10^5$ K to $10^6$ K by the energy released by structure and galaxy formation (e.g. [18,19,24,25]). Evidence for the WHIM from spectroscopic surveys of the low-z IGM has been a major success of FUSE, HST/STIS and HST/COS (e.g. [22,97,93,23]) but the baryon census remains uncertain with 30% still missing ([88,94]).

UV absorption line surveys conducted with much larger effective collecting area than HST hold the promise to probe higher temperature ranges with tracers such as NeVIII and low-metallicity volumes with broad Lyman-α absorbers (e.g. [82,68]). This would fully validate the theories of structure formation that predict the development of a warm-hot phase of the IGM at low redshift.

However, while the general characteristics of the process seem clear, not all the steps are fully understood. One troublesome question has to do with the influence of the environment on galaxies; this has been dubbed the "nature vs. nurture" question in the context of galaxy formation and evolution. It is clear, however, that the IGM is responsible for the formation of the visible galaxies and for their fueling during that time. The IGM itself is modified by galaxy ejecta in the form of galactic winds and by stripping of galaxies of their gas when they pass through a cluster of galaxies.

Starburst galaxies (SBs) are known to drive powerful superwinds. The overpressured $10^{(7-8)}$ K metal enriched plasma that is created by stellar winds and core-collapse SNe sweeps up cooler material and blows it out (or even away). Speeds are of the order of 200-1000 km/s, or even 3000 km/s. SBs can contribute significantly to the mass budget, energetics and metallicity of the IGM. Apart from the metal enrichment [89], superwinds may also create large, ~100 kpc, holes in the IGM [2].

In order to assess the mechanical and chemical feedback from SBs on the IGM, it is necessary to assess the composition and kinematics of the IGM gas, particularly through elements such as O, Ne, Mg, Si, S, Fe; for $T \leq 10^7$ K. In fact, the dominant IGM component is near $\sim 10^5$ K (WHIM) and its mass has been growing with cosmic time to about half the total amount of all the baryons in the local universe.

UV spectroscopic surveys can detect this part of the IGM through Lyα, resonance lines of H, C, O as well as high ionization states of heavier elements, over the last 5-10 Gyr. Particularly CIV, SiIV and OVI are useful probes in this.

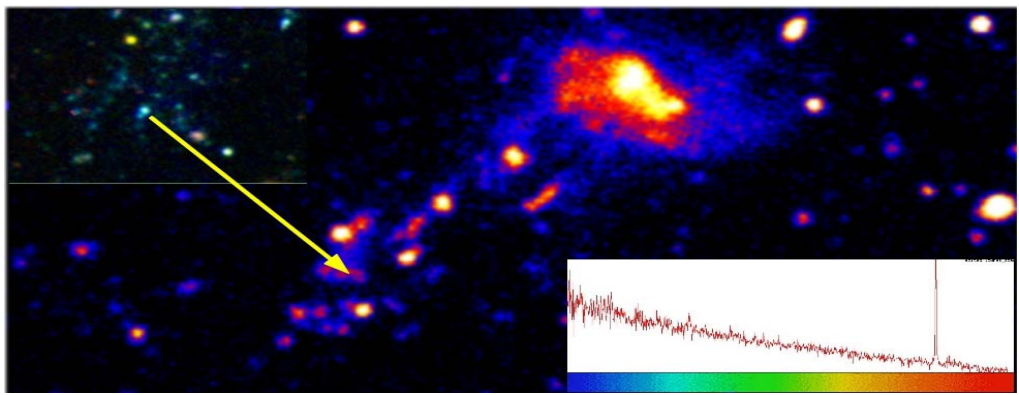

*UV and visible images of IC 3418, a galaxy being stripped in the Virgo Cluster. The arrow points to a supergiant star formed in the stripped material (fig 1 from [76]).*

With the launch of JWST expected to take place within this decade, and with the full ALMA capability to be available even sooner, the study of the very early Universe will receive a very serious experimental boost. It should be noted however that, while the LSS started to emerge at redshifts higher than 3, its evolution continues to the present and what we see now is mainly the product of evolution at z 3. Within the broader cosmological context some unanswered questions are when the first luminous objects formed, what was their nature, and how did they interact with the rest of the Universe. This is also where the questions about re-ionization and metal enrichment arise.

The IGM is enriched with metals by stellar and galactic winds, by ejecta from supernovae and superbubbles, and by winds from active galactic nuclei. The interaction of strong winds with the IGM may partly quench the IGM accretion. Therefore, we witness a feedback process that modifies the IGM dynamics, physical properties and metallicity. However, it is not clear how are the metals distributed in the IGM following their ejection from a galaxy.

Understanding of this process requires studying the sites where metals are produced, which are in massive stars that explode as SNe and in their surrounding nebulosities. Important information will come from detailed observations of low-metallicity local starbursts [e.g. Izw18, one of the most metal-poor galaxies known to date (e.g. [100]) that will enable the investigation of the early evolution of star-forming galaxies, and the origin of high-ionization nebular lines like HeII. Wolf-Rayet (WR) stars are thought to be primarily responsible for the HeII emission. Nevertheless, based on non-rotating evolution models for single stars, a very small number of WR stars are expected to be present in metal-poor systems (see [60] and references therein). In this context, some questions that might be asked are whether the high-excitation lines are really due to WR stars. If so, how can WR stars form in such a low-Z environment? Are the optical spectra adequate to search for WR galaxies? Do some optically-classified WR galaxies host Of-type stars instead of WR stars, and might be misclassified?

The transport of metals into the IGM must be understood and out-flows and in-flows of metals from and into different types of galaxies must be discovered. Metallicity gradients in galaxies, between the disc and the halo, must also be understood. In this context, we mention the understanding of the extended gas distributions of galaxies that can be achieved by studying absorption lines in the lines of sight to distant objects passing through galaxy haloes.

Interactions in dense groups of galaxies, where the dark matter halos have not yet merged into a single entity and the total mass is much less than that of even a poor GC, are best represented by the Hickson compact groups (HCG). These are collections of fewer than ten galaxies in very close proximity to each other. While the original Hickson collection was based on projected surface density, [52] revised the group association via accurate redshift determinations showing that most groups are real. Therefore, the HCGs represent, environments where galaxies are within a few radii and the light crossing times are very short.

At the other extreme one finds galaxies even near the centers of cosmic voids. Such galaxies have had no interactions for (almost) one Hubble time and thus are the best examples of evolution in isolation. A comparison of the two kinds of objects should, therefore, illuminate the case of "nature vs. nurture" in the evolution of galaxies. In particular, one should expect that isolated galaxies in voids should show almost or no star formation, and that their morphological appearance should be very regular.

In general, galaxies in voids tend to be smaller and to exhibit weaker star formation than similar objects in the general field. There is evidence for galaxy formation in voids taking place along DM filaments, since the galaxies appear aligned (e.g., [108,12]). The latter paper presents three galaxies connected within a quasi-linear HI structure, where one of the galaxies VGS_31b exhibits a tail and a ring.

At least one isolated polar ring galaxy (RG) has been identified in a cosmic void [91]. Since the consensus regarding RGs is that these result from a major interaction whereby one galaxy acquires significant amounts of HI from either a nearby companion or directly from the intergalactic space, the conclusion should be that this baryonic mass transfer takes place even in the regions with the lowest galaxy density. Thus neutral IGM can exist in cosmic voids and, when the conditions are suitable, this IGM can convert into luminous stars and galaxies.

The case of RGs is particularly interesting since in many cases these objects appear to have an elliptical or lenticular galaxy surrounded by, or containing, a gaseous and star-forming disc or ring. Since RGs form by accretion of extra-galactic matter, which does not have to follow the kinematic properties of the central object, one finds this type of galaxies to be a good test case for matter orbiting at different orientations within a dark matter halo. Thus the study of RGs offers a way to understand the shape of the DM concentrations.

The GALEX mission has been the only UV wide-field imager yielding significant discoveries in the domain of galaxy evolution. Its images uncovered the extended UV (XUV) discs seen around many spiral galaxies; similar features were detected also around other morphological types of galaxies. The XUV discs are essentially star-forming features. Their location around some objects that had been classified as "red and dead" ellipticals points toward the acquisition of fresh gas either through galaxy-galaxy interactions or directly from the intergalactic space. XUV discs and the RGs may be related phenomena. In order to better characterize them a much more capable instrument is needed than GALEX.

Other intriguing objects newly discovered are "green pea" galaxies (e.g. [17]). Such objects may be similar to but of a higher luminosity than the star-forming very compact "knots" identified near star-forming dwarf galaxies [15] or the "Hα dots" [59]

With the existing telescopes and the giant ones planned form the next decade (TMT, ELT, etc.) it will still not be possible to directly measure the light from individual main-sequence stars in nearby galaxies, since this requires angular resolutions not achievable by ground-based instruments. It will also be possible to analyze in depth blue supergiant stars formed in the stripped material, such as the one recently found [77]

*The requirements for an instrument able to address the issues mentioned above are (i) a large collecting area, (ii) a wide field of view (FOV), (iii) a high spatial resolution, and (iv) the ability to perform medium spectral resolution of point-like objects as well as offering (v) integral field spectroscopy. The large collecting area is necessary to enable the observation of faint sources, since galaxies and individual stars in other galaxies are faint. To probe the extended gaseous haloes via AGN absorptions requires good spectral sensitivity, since the projected spatial density of the AGNs to be used as background sources is sufficient only at faint AGN magnitudes (one QSO per five arcmin at 21 mag.; [50]). The wide FOV is required to allow the sampling of significant numbers of galaxies in e.g. distant galaxy clusters. The high spatial resolution, of about 0".01 or better, is necessary to resolve individual stars in other galaxies and in the cores of globular clusters, as well as to reveal details of structures within galaxies such as HII regions, star clusters, etc. Field spectroscopy is necessary to quantify the stellar populations of resolved galaxies and their immediate neighborhoods.*

2.2 PLANET FORMATION AND THE EMERGENCE OF LIFE
The formation and evolution of exoplanetary systems is essentially the story of the circumstellar gas and dust, initially present in the protostellar environment, and how these are governed by gravity, magnetic fields and the hard radiation from the star. Understanding formation of exoplanets, including the terrestrial ones, and of their atmospheres, calls for a deep study of the life cycle of protostellar discs from their initial conditions to the young planetary discs. UV radiation plays an essential role in this cycle, from disc ionization to chemical processing. It is also the main observational probe of the accretion mechanisms that regulate disc evolution. It also reveals the stellar dynamo that is at the seat of the global magnetic field that plays a significant role in the evolution of exoplanets' atmosphere.

2.2.1 Star-disc interaction: the accretion engine
Accretion onto magnetized structures drives hydro-magnetic processes that power the large scale optical jets and molecular outflows. These engines control the efficiency of accretion and therefore set the final stellar mass by regulating the amount of matter that falls onto the star. During this period, the disc and star form a strongly coupled system. This coupling is eventually broken and the disc dissipates. But the star's magnetic activity continues to influence the evolution of the atmospheres of any nascent planets.

The basic performance of the engine relates with the interaction between the stellar magnetic field and the ionized gas orbiting in the inner border of the disc that generates a toroidal magnetic field. The magnetic pressure from the toroidal field pushes the field lines outwards from the disc rotation axis, inflating and opening them in a butterfly-like pattern and producing a current layer between the star- and disc-dominated regions. Magnetic field dissipation in the current layer also produces high energy radiation and particles above the disc plane. A fraction of the energy released is lost from the system through X rays, EUV, and FUV radiation that heats and ionizes the disc (especially the inner few AU).

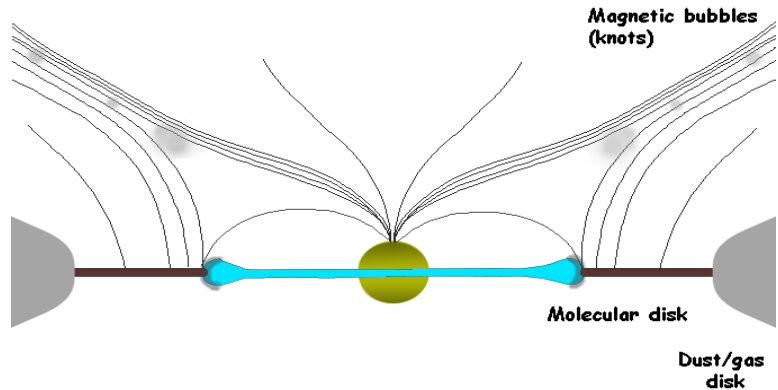

Sketch of the accretion engine. The star acts as a magnetic rotor that interacts with the plasma orbiting around it in Keplerian orbits. The magnetic configuration is outlined from [101] as well as the layer where magnetic reconnection occurs and magnetic bubbles are generated.

The magnetic link between the star and the disc is broken and reconnected, continually ejecting blobs of plasma: the knots in the optical jets ([49,101]). The opening angle of the current layer, and its extent, depends on the stellar and disc fields, the accretion rate and the ratio of the inner disc radius and stellar rotation frequencies. The jet radiates in UV intercombination SiIII],CIII] [43,44] at 1892Å and 1908Å that are optically thin and can be used to probe the warm base and to study the jet collimation mechanisms. Moreover, UV instrumentation provides the best possible spatial resolution for studying any jet rotation at the base [21]. The first detection of the hot solar-like winds in T Tauri stars (TTSs) has been achieved using their NV emission [46,48].

The star-disc interaction also distorts and powers the stellar magnetosphere, leading to the formation of structures extending several stellar radii [84,85] that dominate the UV radiation from the TTSs. UV radiation during the T Tauri phase is typically 50 times stronger than during the main sequence evolution of solar-mass stars [57,107,47,4]. Both line strength and broadening decreases as the stars approach the main sequence [4,48] confirming that the UV excess is dominated by the extended magnetospheric emission. The correlations between the UV magnetospheric tracers seem to extend from TTSs to brown dwarfs [47] providing a unique tool to study the physics and evolution of the engine for a broad range of masses. Unfortunately, the UV spectrum of just one brown dwarf [35] has been obtained because of the lack of sensitivity.

Many uncertainties remain about how star-disc interaction self-regulates and also regarding the dependence of the engine details on initial conditions e.g. the effective gravity of the star, the role of stellar radiation and magnetic field on the engine performance and the role of the ionizing radiation produced by the engine on the evolution of the disc. Despite the modeling advance so far, the real properties of the engine itself are poorly known given the lack of observations to constrain the modeling. Very important questions still open are: *How does the accretion flow proceed from the disc to the star? Is there any preferred accretion geometry as, for instance, funnel flows? What is the temperature distribution emerging from the accretion shock? What role do disc instabilities play in the whole accretion/outflow process? What are the dominant processes involved in wind acceleration? What are the relevant time-scales for mass ejection? How does this high energy environment affect the chemical properties of the disc and planetary building? Whether and how this mechanism works when radiation pressure becomes significant as for Herbig Ae/Be stars? How does the stellar magnetosphere evolve?*

A single spectrum in the UV range contains information about all the physical components - atmosphere, magnetosphere, outflows (Solar-like winds, jets), accretion flow, inner disc structure, residual gas in the young planetary system - and their evolution into exoplanetary systems. At least, a factor of 10 improvement on sensitivity is required (reaching $10^{-17}$ erg/s/cm$^2$/A with R$\approx$20,000) to observe the faintest components of the engine in a sample large enough to answer the above questions. This represents a factor of ten improvement over current facilities (HST).

### 2.2.2 The evolution of the gaseous component in the disc

The lifetime, spatial distribution, and composition of gas and dust of young (age < 30 Myr) circumstellar discs are important properties for understanding the formation and evolution of extrasolar planetary systems. Disc gas regulates planetary migration [102,5,96] and the migration timescale is sensitive to the specifics of the disc surface density distribution and dissipation timescale [6]. Moreover, the formation of giant planet cores and their accretion of gaseous envelopes occurs on timescales similar to the lifetimes of the discs around T Tauri and Herbig Ae/Be stars ($10^6 - 10^7$ yr).

UV radiation from the engine is very efficient at etching the disc surface through photoevaporative flows [53,3] and determines the lifetime of the gaseous component of discs. The dust disc clearing timescale is expected to be 2-4 Myr [51], however recent results indicate that inner molecular discs can persist to ages ~10 Myr in Classical TTSs [86,56,37]. Fluorescent $H_2$ spectra in the 912 – 1650 Å bandpass are sensitive to probe gas column densities <$10^{-6}$ g cm$^{-2}$, making them the most sensitive tracer of tenuous gas in the protoplanetary environment. While mid-IR CO spectra or other traditional accretion diagnostics suggest that the inner gas disc has dissipated, far-UV $H_2$ and CO observations offer unambiguous evidence for the presence of an inner (r<10 AU) molecular disc [36,38]. The penetration of UV photons from the engine into the dusty disc and the spectral distribution of the FUV radiation field have a strong influence on the chemical abundances of the disc [11]. FUV radiation photo dissociating $H_2$ and CO controls the chemistry of the external layers of the disc.

Spectroscopic observations of volatiles released by dust, planetesimals and comets provide an extremely powerful tool for determining the relative abundances of the vaporizing species and studying the photochemical and physical processes acting in the inner parts of the planetary discs [98,65,66].

### 2.2.3 Influence of UV emission on planets

The "habitability" of planets depends not only on the thermal effect of the central star but also on its magnetic activity [79], which strongly influences the chemical (and possibly biochemical) processes at the surface of the planet. Stellar UV and EUV emission is an important contributor to planetary atmosphere evaporation by photo-dissociation and is also responsible for damage to the biochemical structures necessary for a biological activity [87]. UV and EUV fluxes originate in hot plasma in the stellar atmosphere. The required temperature (from some tens of thousands to million Kelvin) is reached by the non-thermal heating of the atmosphere caused by the action of the stellar dynamo (see Sect. 2.4).

Solar-like stars show high variability in this energy band [76, 38, 40, 78]. Contrary to the corona, where the plasma is optically thin, the transition region and the chromosphere are optically thick releasing high UV fluxes. During their lives, stars lose angular momentum through torquing by magnetized stellar winds. As a result, high energy emission from solar-like stars decreases with age. As their magnetic activity intensity is connected with their rotation rate [90,80], UV emission also follows this trend [71,73]. As a consequence, planetary atmospheres receive higher UV radiation during the first stages of the planet's life. UV and EUV radiation strongly influence planetary atmospheres (e.g., [64]). A spectacular case has been revealed through planetary transit studies: the star induces massive (and irregular) planetary material ejection because of the close planet orbit (0.01 AU), the planet shows a comet-like tail, responsible for the observed variations of transit depth [81]. This example, together with the influence of magnetic activity on planetary habitability, shows how necessary it is to understand stellar dynamos to describe the orbiting planets; it is in fact insufficient to extrapolate EUV and UV fluxes from X-rays, being demonstrated that correlation between emission in the two spectral regions is not universal, but strongly dependent on the activity level of the star. For example, [92] show no correlation between UV and X-ray flux for slowly-rotating M dwarfs. Moreover, planet detection and characterization rely directly on knowledge of its stellar host (see Sect. 2.3).

### 2.2.4 Characterization of exoplanet atmospheres

Understanding the physical processes in exoplanetary atmospheres requires studying a large sample of systems. Many planets have been discovered through transits, for example by the Corot and Kepler missions and will be discovered with GAIA or CHEOPS. The most recent estimate [28] is that GAIA will discover some hundred transiting Hot Jupiters (P <5 d) to G <14, and a few thousand to G <16. These will be prime targets for detecting atmospheric constituents through absorption spectroscopy, thereby characterizing the chemical and physical properties of the atmosphere. The first detections and observation of the atmospheres of transiting extrasolar planets (HD209458b and HD189733b) have been made through UV-visible spectroscopy [99,69,67].

The observation of UV and visible absorption when a planet transits its parent star is a very powerful diagnostic technique. In fact, this method is the most effective way to detect Earth-like planets because of the strong absorption of stellar UV photons by the ozone molecule in the planetary atmosphere (see [45]). Observation of biomarkers in the high atmosphere of Earth shows that these atoms and ions are present at very high altitude (even several hundreds of kilometers) causing large absorption depths. *Electronic molecular transitions*, pumped by UV photons, *are several orders of magnitude stronger than the vibrational or rotational transitions observed in the infrared or radio* range provided the stellar UV radiation field is strong enough. For a typical life-supporting terrestrial planet, the ozone layer is optically thick to grazing incident UV radiation to an altitude of about 60 kilometers. With a telescope 50 times as sensitive as HST/STIS, ozone can be detected in earth-like planets orbiting stars brighter than V~10 (easily discovered by GAIA). This magnitude corresponds to a star at a distance d~50 pc for the latest type stars considered (K dwarf stars) and more than ~500 pc for the earliest stars (F dwarf stars).

2.2.4a Upper atmosphere emissions
Known planetary atmospheres radiate various non-LTE emission lines in the EUV-FUV. Among them, the giant planets are the richest and strongest emitters (Ly$\alpha$, $H_2$ Lyman and Werner bands, OI1300Å, etc…). These emissions directly originate from the stellar UV flux and the precipitation of energetic particles into the upper atmosphere. Transposed to exoplanets, the detection of such direct emissions would provide crucial information on the atmospheric composition and dynamics (e.g. atmospheric escape) or the processes responsible for energy transport and deposition. Fluid and thermodynamical states could be inferred from these emissions, as homo-bases and exo-bases altitudes, temperatures, variability etc.

The main difficulty with this approach is to discern the planet emission lines from the parent star emission. [34] reported a possible observational detection of $H_2$ emissions at 1580 Å for HD209458b. There are also tantalizing hints of photo-excited molecular emission from planets orbiting M-dwarfs [38,40]. In parallel, theoretical work [74] has calculated contrast up to $10^{-2}$ for Ly$\alpha$ emission of the same planets, which can theoretically be achieved for the closest planets (d< 20pc).

2.2.5 The bulk composition of exo-planets
Measuring the bulk composition of extrasolar planetary material can be done from high-resolution UV spectroscopy of debris-polluted white dwarfs [41]. Most known planet host stars will evolve into WDs, and many of their planets will survive (e.g. Mars and beyond in the solar system). The strong surface gravity of WDs causes metals to sink out of the atmosphere on time-scales much shorter than their cooling ages, leading unavoidably to pristine H/He atmospheres. Debris discs form from the tidal disruption of asteroids [58] or Kuiper belt-like objects [13], are stirred up by left-over planets [26], and can subsequently be accreted onto the white dwarf, imprinting their abundance pattern into its atmosphere.

*The requirements for an instrument able to address the issues mentioned above are (i) a large collecting area, (ii) a wide field of view (FOV), (iii) a high spatial resolution, and (iv) the ability to perform high dispersion (D=20,000-50,000) spectroscopy in the full 920-3200 A range, as well as offering (v) integral field spectroscopy with moderate dispersion (D=3000) over large fields of view (10'x10'). The large collecting area is necessary to enable the observation of faint M stars and brown dwarf. Sensitivities of $10^{-17}$ erg s $cm^{-2}$ $A^{-1}$ are required to obtain good S/N profiles of the target lines. Spectral coverage in the 912 – 1150 A bandpass is needed as the bulk of the warm/cold $H_2$ gas is only observable at l < 1120 A (via the Lyman and Werner (v' - 0) band systems). Spectropolarimetry would permit following the evolution of the dusty plasma in the circumstellar environment. Dynamical ranges above 100 and resolutions larger than 100,000 are required to separate stellar and planets contribution.*

2.3 THE SOLAR SYSTEM

The various bodies in the solar system provide different, complementary pieces of the puzzle that is Solar System formation, the planets having formed by coalescence of planetesimals, of which evidence remains in the form of asteroids and Trans-Neptunian Objects (TNOs). With its access to thousands of resonance lines and fluorescence transitions, the discovery potential of the proposed UV-visible observatory is vast, covering planetary atmospheres and magnetospheres, planetary surfaces and rings, comets, TNOs and other small solar system bodies, and interplanetary material [16]. Because UV photons interact strongly with matter, UV observations are excellent to determine the composition and structure in low-density regions of the solar system, where plasmas and atmospheres interact, and UV irradiation drives solid-phase and gaseous chemistry, the latter of which dominates the structure of planetary atmospheres above the tropopause. Further, the high spatial resolution afforded in the UV enables exploration of the entire population of solar system bodies as spatially-resolved targets. For example, an 8m diffraction-limited UV telescope will provide spatial resolution of ~8 and ~55 km at Jupiter's and Neptune's distances, respectively, at 1000 Å. This compares extremely favorably with in situ spacecraft observations; at closest approach to Jupiter Juno's UV spectrometer will provide pseudoimages with ≈60 km spatial resolution. A non-exhaustive list of science goals is as follows:

2.3.1 Atmospheres and magnetospheres

This UV observatory will be capable of revolutionizing the study of the dynamics and composition of planetary and satellite atmospheres, particularly at the poorly understood planets of Uranus and Neptune. Observations of the abundance and distribution of species such as H, $H_2$, $CO_2$, CO, and $H_2O$, along with many organics and aerosols will be possible. These provide essential insight into source and loss processes, volcanism, aeronomy, atmospheric circulation and long term evolution of planetary atmospheres. These phenomena all connect to wider issues, such as historical and present habitability, terrestrial anthropogenic climate change, and the nature of the presolar nebula. Cyrovolcanic and dust plumes, like those observed in situ at Enceladus are potentially within the reach of this project along with detailed studies of other small bodies -- planetary satellites and trans-Neptunian objects (TNOs). Through sensitive, high resolution imaging of planetary and satellite auroral emissions, this UV observatory will also impart a detailed understanding of all of the planets' magnetospheres, revealing the internal magnetic fields and thus internal structure and formation, along with information as to how energy and matter flow through the solar system. Short of sending dedicated spacecraft, this UV telescope is the only way to investigate the magnetospheres of the ice giants Uranus and Neptune. Jupiter's magnetosphere in particular acts as a readily-observable analogue for more distant astrophysical bodies such as exoplanets, brown dwarfs and pulsars. Importantly, this observatory will not duplicate but instead perfectly complement ESA's L1 mission JUICE.

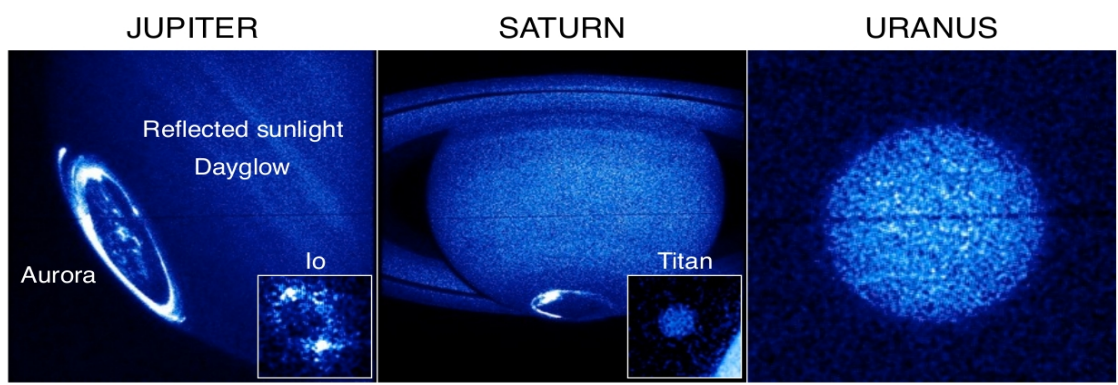

A collage of some solar system UV targets. From left to right, these include the atmospheres, auroras and airglow of Jupiter, Io, Saturn, Titan, and Uranus.

2.3.2 Surfaces and rings

Surface spectroscopy and imaging will provide information on the ice and non-ice condensible (e.g. organic) components of surface layers, revealing interior processes and surface-atmosphere interactions, with important implications for habitability at e.g. Europa. Long term observations of albedo maps will allow the study of seasonal effects. Observations of H, ice, organics and other minor species in planetary ring systems will reveal their composition, formation- and life-times.

2.3.3 Small bodies

The main water products in cometary comas, H, O, and OH, along with the CO Cameron bands, can be uniquely observed at high spatial resolution in the UV. These observations, along with high sensitivity detections of C, S, N, D/H, and rare gases beyond the snow line, unveil the temperature and density of the pre-solar nebula in which the comets formed. Further, observations of O+ and CO+ will probe the interaction of comets with the solar wind. A major objective will be to detect comet-like activity in TNOs, Main Belt asteroids, Trojans and Centaurs, testing models of thermal evolution at large heliocentric distances. High angular resolution albedo maps and observations of binarity of thousands of Main Belt asteroids will provide information on their composition, and thus source material, while many tens of bright TNOs will be fully characterized by these UV observations.

*Solar system science is largely concerned with temporal phenomena, with characteristic timescales varying from seconds to years. A highly-sensitive UV observatory placed at L2 will allow high temporal resolution data to be obtained over long, unocculted intervals, providing a revolutionary advancement over previous Earth-based and, in situ, UV platforms. With its ability to observe many solar system targets, this observatory will uniquely provide the holistic approach required to unravel the story of the solar system.*

2.4 THE INTERSTELLAR MEDIUM (ISM)

The UV domain is very rich in information for absorption studies of the ISM, for both its gas and dust phases. For gas studies, the UV domain is fundamental because it includes most important gas phase transitions: from hydrogen in its different forms HI, DI, $H_2$, HD and from the important heavy elements like C, N, O, Si, Fe, Mg in their dominant ionization stages in neutral regions as well as in ionized and highly ionized media. This is the only way to obtain a direct census of the hydrogen fraction of the Galaxy. All of its forms are accessible in multiple resonance transitions and this provides the key link to the radio and infrared studies of the dust and diffuse gas. These diagnostic lines therefore probe gas from below 100 K to above $10^6$ K and allow to derive precise abundances, ionization, cosmic rays flux and physical parameters as temperature, pressure and density for a wide variety of media, from coronal gas to diffuse neutral and ionized clouds, and translucent or molecular clouds. In the case of the dust, the UV provides a powerful tool with the extinction curves, which presents large variations in the UV and include the largest dust feature at 2175 A.

Extensive studies of the Galactic ISM have been performed with the previous UV facilities, in particular GHRS and STIS on board HST for wavelengths down to 1150 A, and Copernicus and FUSE for the far-UV range. *Significant progress on this field will be achieved provided the future instrument has the potential to: (1) increase the sensitivity relative to past or existing spectroscopic facilities in order to extend the ISM studies to other galaxy discs and halos; (2) observe with the same instrument as many different species as possible to probe gas at different temperatures and densities and cover the multiphase aspect of the interstellar medium in galaxies, from the cold phases probed e.g. with CI to the hot phases responsible for the ubiquitous OVI; (3) get simultaneous access to the Lyman series of atomic hydrogen and to the Lyman and Werner bands of molecular hydrogen to derive abundances, gas-to-dust ratios, and molecular fractions; (4) get high resolution of at least 100 000 in the observations of H2 (and other molecules CO, HD, N2...), to resolve the velocity structure of the molecular gas, measure line widths for the various J-levels to study turbulence motion and excitation processes in molecular and translucent clouds; (5) extend the observations to high Av targets in the Milky Way to extend the range of molecular fractions and study depletion and fractionation in denser molecular cores; (6) produce high signal to noise ratios in excess of 100 to measure weak lines like those produced by the very local ISM, from both low ionization species and high ionization species like OVI and CIV, to describe the detailed structure of the local cloud(s) and characterize the nature of the hot gas in the Local Bubble; and (7) observe large samples of targets in a wide range of environments in the Galaxy and other galaxies and couple dust and gas studies of the same sightlines.*

## 2.5 STELLAR PHYSICS

The UV domain is crucial in stellar physics, all stellar studies benefit from access to the UV range, and some are actually impossible without it. The intrinsic spectral distribution of hot stars peaks and the resonances lines of many species, prone to non-LTE effects, probe the highest photospheric layers, or winds (CIV, NV, etc...), or non-radiative heating in chromospheres in cool stars. Another advantage of UV observations is the extreme sensitivity of the Planck function to the presence of small amounts of hot gas in dominantly cool environments. This allows the detection and monitoring of various phenomena otherwise difficult to observe: magnetic activity, chromospheric heating, corona, starspots on cool stars, and intrinsically faint, but hot, companions of cool stars. The UV domain is also where Sun-like stars exhibit their hostility (or not) towards Earth-like life (see Sects. 2.2 and 2.3), population III stars must have shone the brightest, accretion processes convert much kinetic energy into radiation, the "Fe curtain" features respond to changes in local irradiation, flares produce emission, etc. Moreover, many light scattering and polarizing processes are stronger at UV wavelengths. Organized global magnetic fields in stars interact with their wind and environment, modify their structure and surface abundances, and contribute to the transport of angular momentum. With spectropolarimetry, one can address with unprecedented detail these important issues (from stellar magnetic fields to surface inhomogeneities, surface differential rotation to activity cycles and magnetic braking, from microscopic diffusion to turbulence, convection and circulation in stellar interiors, from abundances and pulsations in stellar atmospheres to stellar winds and accretion discs, from the early phases of stellar formation to the late stages of stellar evolution, from extended circumstellar environments to distant interstellar medium).

Measuring polarization directly in the UV wind-sensitive lines has never been attempted and would be a great leap forward in studying the nonsphericity and magnetic effects thought to be present in stellar outflows. The sampling by a space-based instrument will yield continuous time-series with short-cadence measurements. Such time-series document phenomena on stars that can be impulsive (flares, infall), periodic (pulsations, rotational migration of spots, co-rotating clouds), quasi-periodic (evolution of blobs from hot winds), or gradual (evolution of spots). Such an instrument will thus provide a very powerful and unique tool to study most aspects of stellar physics in general. In particular, it will answer the following long-standing questions, as well as new ones: *What is the incidence of magnetic fields? In which conditions does a dynamo magnetic field develop? Magnetic field dynamics and geometry: how do they affect all aspects of stellar structure and evolution? What are the properties of wind and mass loss? How does a stellar magnetic field influence mass loss, in particular what is responsible for wind clumping, the formation of a circumstellar disc or clouds, and flares? Under what conditions do OB stars become Be stars? How do their discs form, and dissipate again? What causes Luminous Blue Variable outbursts? What happens when a star reaches critical rotational velocity? What is the origin of γ Cas stars behavior? How does binarity affect stellar structure and evolution? What are the properties of the galactic white dwarf population, and what can they teach us regarding fundamental physics? How does accretion occur in X-ray binaries? What is the interplay between mass transfer and radiation pressure? Is UV/visible variability related to X-ray emission? What evolutionary channels lead to type Ia supernovae?* These questions will be answered by studying various types of stars, especially:

### 2.5.1 Hot stars:

Early-type (OB) stars dominate the ecology of the universe as driving agents, through their luminosities and mechanical inputs (e.g. winds, supernova explosions, novae). For that reason they all display, at least at some moment of their life time, strong variability on a wide range of timescales. This concerns, for example, Of?p stars which have very specific spectral characteristics related to their magnetic field, Be stars which are very rapidly rotating and develop decretion circumstellar disc, γ Cas stars which emit unexplained variable X-ray flux, Bp stars which host very strong fossil magnetic fields, Herbig Be stars which are the precursor of main sequence Ap/Bp stars, β Cep and Slowly Pulsating B (SPB) which pulsate, B[e] with dust and of course O stars, as well as massive binaries such as the Be X-ray binaries and those that harbor O-type subdwarf companions, etc. They are also unique targets for the study of stellar magnetospheres. Their strong, radiatively-driven winds couple to magnetic fields, generating complex and dynamic magnetospheric structures [7,27], and enhancing the shedding of rotational angular momentum via magnetic braking [106,75]. As the evolution of massive stars is particularly sensitive to rotation and mass loss [20,72], the presence of even a relatively weak magnetic field can profoundly influence the evolution of massive stars and their feedback effects, such as mechanical energy deposition in the interstellar medium (ISM) and supernova explosions (e.g. [30]). Stellar winds from hot massive stars can be structured also on small-scales by the intrinsic "line-driven instability" (LDI). The presence and interactions between density structures on both these scales is poorly understood, and may compromise the reliability of measurements of the properties of the outflows. Spectral diagnostics such as UV resonance and visible recombination lines have different dependencies on density, and will provide crucial constraints for the further development of dynamical hot star wind models, as well as for how the resulting wind structures affect derived quantities such as mass loss and rotation, which are essential inputs for corresponding models of stellar evolution and feedback. Indeed, if mass-loss is poorly constrained, the evolution of massive stars, their fate and feedback can be completely misunderstood. Although clumping appears to be a universal feature of line-driven winds, it is not known how the LDI interacts with other processes that structure the wind. Some key questions concern possible inhibition of the lateral fragmentation of clumps, the effects on the structure within the closed field loops (the so-called "dynamical magnetosphere"), and how these different behaviors alter the interpretation of spectral diagnostics, in particular the determination of mass-loss rates. EUVO can address these issues by providing time-series at high spectroscopic resolution for OB stars.

2.5.2 Binary stars
Although their evolution is often treated in isolation, about half of all stars in the Galaxy are members of binary or multiple systems. In many cases this is not important but when it is, and that is not rare, the effects can be best studied in the UV. Magnetic fields play a central role, as they strongly affect, and are strongly affected by, the transfer of energy, mass and angular momentum between the components. However, the interplay between stellar magnetospheres and binarity are poorly understood, both from the observational and theoretical side. In higher-mass stars (> 1.5 $M_o$) the incidence of magnetic stars in binary systems provides a basic constraint on the origin of the fields, assumed to be fossil, and on whether such strong magnetic fields suppress binary formation. In low-mass stars, tidal interactions are expected to induce large-scale 3D shear and/or helical circulation in stellar interiors that can significantly perturb the stellar dynamo. Similar flows may also influence the fossil magnetic fields of higher-mass stars. Magnetically driven winds/outflows in cool and hot close binary systems have long been suspected to be responsible for their orbital evolution, while magnetospheric interactions have been proposed to enhance stellar activity. The ultraviolet is of central importance for studying the complex phenomenon of stellar magnetism under the influence of the physical processes and interactions in close binary systems.

2.5.3 White dwarfs
>95% of all stars will eventually evolve into white dwarfs, and their study is fundamental to a complete understanding of stellar evolution, with implications into galaxy evolution (through the initial-to-final mass relation) and no picture of stellar or galactic evolution can be complete without them. Detailed photospheric abundance measurements are only possible in the FUV (e.g. [8]). However, because they are intrinsically faint, only a handful of white dwarfs have been thoroughly studied with HST and FUSE. The large effective area of EUVO is indispensable to observe a representative sample of a few hundred stars spanning a wide range of ages, masses, and core/photospheric compositions. Narrow metal lines in the FUV spectra of white dwarfs are a powerful diagnostic for a range of physics, including the very sensitive search for low magnetic fields, or the detection of the coupling between scalar fields and gravitational potential [31].

2.5.4 Compact binaries
Binaries containing a white dwarf (WD), neutron star (NS), or black hole (BH) represent some of the most exotic objects in the Universe, and are ideal laboratories to study accretion and outflow processes [69], and provide insight on matter under extreme conditions. Which evolutionary path close compact binaries follow is a key but still unresolved problem. Dynamical evolution of the binary system proceeds through angular momentum loss and tidal coupling. How this occurs has implications for a wide range of other open questions. One of the fundamental questions is the nature of SN Ia progenitors, the standard candles tracing the existence of dark energy. Whether SN Ia descend from single degenerate or double degenerate binaries, or both, is still controversial. Mass accretion makes the evolution of WDs in such compact systems essentially different than isolated stars (e.g. [104]). The study of photospheric emission of these degenerates, as well as the physical and chemical conditions of the accretion flow are crucial to trace the mass transfer/accretion history and the effects on the evolution. This can be only achieved in the UV range through high resolution FUV spectroscopy of statistically significant samples of accreting

WDs spanning a wide range in stellar and binary parameters including magnetic field strengths. In this respect, the high incidence of magnetic accreting WDs compared to single WDs have lead to different but still debated proposals [95].

A large UV-visible mission is also critical to comprehending accretion disc physics in X-ray binaries and of how the disc reacts to changes in the mass transfer rate or how instabilities are driven. Rapid response to transient events, such as Novae, Dwarf Novae and X-ray transients, is of key importance to unveil changes in the physical conditions of accreted or ejected flows through the outburst evolution of the UV emission lines and continuum. Only very recently the onset of jets has also been identified in WDs accreting at high rate [61,62], suggesting that disc/jet coupling mechanism is ubiquitous in all types of binaries (BHs, NS and WDs). Also, the ratio of UV and X-ray luminosities is recently recognized as important discriminator between NS and BH binaries [52] and that FUV continuum could be affected by synchrotron emission from the jet. Furthermore, in the case of wind accretion, the detailed analysis of the wind structure and variability will improve our knowledge both in terms of how accretion takes place, and how mass loss rate is affected by photo-ionization from X-rays (see e.g. the theoretical work [49] and observational work [56] (FUSE), and [25] (HST)). When such L2 mission will be operational, GAIA will have provided accurate distances for a thousand of cataclysmic variables (CVs) and tens of X-ray binaries allowing tight observational constrains to theories of accretion and evolution. In addition UV imaging with high-temporal capabilities can efficiently allow identifying exotic binaries in Globular Clusters such as ultracompact LMXBs (UCBs). These are expected to be abundant in GC cores HST has so far, only identified three.

2.5.6 Supernovae
Ultraviolet spectrophotometry of supernovae (SNe) is an important tool to study the explosion physics and environments of SNe. However, even after 25 years of efforts, only few high-quality ultraviolet (UV) data are available – only few objects per main SN type (Ia, Ib/c, II) – that allow a characterization of the UV properties of SNe. EUVO could be of paramount importance to improve the current situation. The high-quality data will provide much needed information on the explosion physics and environments of SNe, such as a detailed characterization of the metal line blanketing, metallicity of the SN ejecta, degree of mixing of newly synthesized elements, as well as the possible interaction of the SN ejecta with material in the environment of SNe. The utility of SNe Ia as cosmological probes depends on the degree of our understanding of SN Ia physics and various systematic effects such as cosmic chemical evolution. We now know that some "twin" Type Ia supernovae which have extremely similar optical light curves and spectra, they do have different ultraviolet continua [33]. This difference in UV continua was inferred to be the result of significantly different progenitor metallicities. Early-time UV spectrophotometry with UVO of nearby SNe Ia will be crucial for understanding the detailed physics of the explosions, determining if SNe Ia have evolved (and by how much) over cosmic time, and fully utilizing the large samples high-redshift SNe Ia for precision cosmology measurements.

*To address these issues one requires (i) a large collecting area, (ii) a wide field of view (FOV), (iii) a high spatial resolution, and (iv) high spectral resolution as well as (v) integral field spectroscopy. Furthermore, the progress achieved in stellar physics thanks to simultaneous UV and visible high-resolution spectropolarimetry will revolutionize our view of stars of all types and age but it requires an increase in sensitivity with respect to HST by a factor of 50-100 to reach the S/N required for the observation of most of the targets. High-resolution spectropolarimetry will make feasible to produce 3D maps of stars and their environment, and understand the impact of various physical processes on the life of stars. These results will have an important impact on many other domains of astrophysics as well, such as planetary science or galactic evolution.*

2.5 OTHER RELEVANT PROBLEMS
Fundamental physics – testing the variation of the fine structure constant at z:
The absorption of light along the line of sight to quasars by intervening gas clouds provides a mean to measure the variation of the fine structure constant across the Universe. The fine structure constant, $\alpha=e^2/\hbar c$, is the parameter that governs the strength of electromagnetism; it couples the electromagnetic field to all charged particles in nature. The separation in wavelengths among transitions in the same multiplet depends on α. For instance, alkali-doublet (AD) type transitions, the separation is proportional to $\alpha^2$. The Many Multiplets (MM) method compares many transitions from different multiplets, from different atoms/ions yielding uncertainties as low as Δα/α≈ $10^{-6}$ [29,104,61]. Indications of variations of $\alpha$ with redshift and location in space have been obtained from measurements of the resonance UV multiplets of Fe II, Ni II, Cr II, Zn II and Mg II [106]. These multiplets are observed in the visible range for redshifts z>1.6 henceforth, the current measurements have been obtained with 10-m class ground based telescopes. As shown in the figure, the measurements are very uncertain, especially because of the atmospheric effects (refraction index, sky lines, [61]). Measurements from space would be much more accurate provided that stable high resolution spectroscopy and a high collecting surface to reach z=2 is provided. Also, space opens the possibility of using stronger multiplets like the Lyman series of hydrogen.

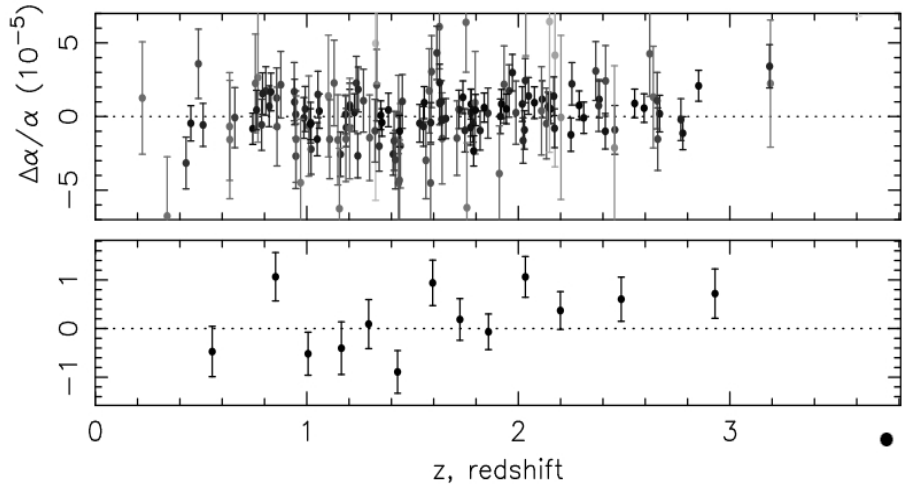

Δα/α measured with VLT/UVES. The bottom panel shows binned values of Δα/α where approximately 12 points contribute to each bin from [61]

## 3. PROPOSED INSTRUMENTATION

The Hubble Space Telescope (HST) and its suite of instruments have inevitably set the standard for UV and visible astronomy from space. Conceived as a general-purpose observatory, it has been able to study a wide variety of scientific goals by providing diffraction-limited imaging and spectroscopy from low to high resolution across a wavelength range from the far ultraviolet to the near infra red. In its current configuration, the telescope is at the peak of its capability and continues to deliver exciting new discoveries. Therefore, any new facility proposed needs to reach well beyond the capabilities of HST. The scientific goals proposed for a new UV/visible observatory require dramatic increases in sensitivity, which must be driven by increases in collecting area and efficiency of the various instruments. A factor 50-100 increase in overall throughput is required to achieve the scientific advances proposed in this white paper. This will be achieved by combination of increased geometric area, through a much larger mirror than HST coupled with advances in reflective coatings, reduced number of reflections through a judicious optical design, enhance detector efficiency, and spectrograph design. However, the required increase in geometric area is likely to be larger than can be provided by a conventional monolithic design contained in the currently available launcher fairings. The baseline telescope design is as follows:
- 8m deployable segmented optical design
- wide-field diffraction limited imaging detector system, with angular resolution 0.01 arcsec
- UV spectrograph with low/medium to high resolution echelle capability, R=20,000-100,000; 900-2000Å and 2000-4000Å, including a long slit or multi-slits
- Spectropolarimetry, R=100,000 (1000Å-7000Å)
- Integral field spectroscopy, R=500-1000
- Detectors should have photon counting capability

The following sections discuss these elements in more detail and outline required technology developments.

## 3.2 Telescope Design

Conventional telescope designs have traditionally been used on space systems. HST is a Ritchey-Chretien design with a 2.4m aperture while the largest telescope launched into space so far (by ESA) is the 3.5m mirror of the Herschel mission. The simplest configuration to implement, a monolithic mirror with a fixed structure, is limited by the available launcher systems and fairing dimensions. For example, the Ariane 5 fairing limits us to a 4m aperture telescope yielding an increase in collecting area of ~2.8 compared to HST. Therefore, achieving the necessary factor 50-120 improvement in effective area a large increase in the telescope mirror dimensions is required coupled with significant enhancements in other areas.

Further enhancements in collecting area require technological development beyond the current state of the art. There are three main possibilities:
1. Enhanced launcher configuration with larger fairing – there are currently no relevant planned upgrades.
2. Off-axis elliptical Mirror design to the largest size acceptable by the fairing – gives only a further factor 2 increase in area, but requires a deployable structure for the secondary mirror.
3. Deployable folded mirror design – JWST has 6.5 m aperture (7.5 x HST area) and uses an Ariane 5 launch, but the system will need enhancements to fulfill the needs of a UV/visible mission. However, a folded mirror system is the only option that can deliver the necessary collecting area, requiring an aperture ~8m.

Some of the necessary technological developments are being addressed in the ESA future technology planning, as well as by NASA technology calls.

## 3.3 Optical Coatings

The efficiency of reflective optical coatings is of crucial importance to the efficiency of UV/visible telescopes, particularly for the shorter wavelengths where reflectivity can be low. Complex optical systems including primary/secondary mirrors, pick-off mirrors and gratings compound the problem to the power of the number of reflections in the system. For example, a reflectivity of 75% yields a net efficiency of 18% after 3 bounces. If the reflectivity could be improved to 90%, the net efficiency would be 73%, a significant, factor 4, improvement. The standard coating as used in HST is $MgF_2$ overcoated Al. However, its performance becomes problematic at the shorted FUV wavelengths (below 1150 Å). Alternatives such as SiC and LiF, have considerably better short wavelength performance but are typically only ~60% at the longer wavelengths. This is an area where technical development is already underway, with some promising results, but where further work is required. For example, a very thin $MgF_2$ coating on Al using Atomic Layer Deposition (ALD) techniques can offer efficiencies >~50% below 115nm while retaining high (90%) performance at the longer wavelengths.

## 3.4 Detector Systems

Separate detector systems will be required for the imaging and spectroscopic elements of the mission payload. Microchannel Plate (MCP) detectors have been the device of choice in the FUV for all recent missions, while CCDs have been used for NUV and visible bands. Each has advantages and disadvantages. CCDs are integrating devices, require cooling to reduce noise and have limited QE at the shorter wavelengths. MCPs are photon counting, have better QE at short wavelengths but have limited count rate capability and suffer gain sag over time. An ideal detector system would combine the best features of both devices - high QE, good dynamic range, long-term stability and low noise. Future detector developments may also include ICCD, sCMOS or ICMOS devices.

Back illumination, delta doping and AR-coating can improve the UV QE of CCDs (>50% @ 1250Å – L3CCD, JPL). The first devices are now being tested on sub-orbital missions. CMOS devices have similar QE performance to CCDs, and their integrated nature, with all ancillary electronics inbuilt in the device produces a compact, lower mass, lower power detector with high radiation tolerance. Radiation damage caused by cosmic rays in CCDs can cause increased dark noise, image artifacts and hot pixels. "Low Light Level" CCDs have electron multiplication, which can make their performance close to photon counting with suitable cooling (<-100C), but have higher stabilization requirements and other operating issues including ageing effects. In the long term, CMOS devices are likely to replace CCDs as general sensors. Therefore, it is important that development work is focused on the UV performance of the former.

The performance of microchannel plate detectors is still improving; advances in MCP technology using coatings applied to borosilicate glass MCPs with atomic layer deposition (ALD) can provide higher MCP gain and detection efficiency and significantly extend detector lifetime, while readout developments employing lower noise techniques (such as capacitive division) in conjunction with adaptive digital filtering enable higher dynamic range to be achieved. ITAR restrictions on ALD coated MCPs is currently imposed only for MCPs of pore size 5 microns or smaller. A UK collaboration has recently been funded to investigate application of ALD to photon counting detector techniques.

The electronic readouts are a key element of the performance of MCP detectors. For example, a capacitive division image readout (C-DIR) with pulse digitization and scene-dependent adaptive digital filtering algorithms to combine very high resolution MCP-limited imaging at moderate count rates whilst extending the dynamic range of the detector to several Mcount/s. The combination of the low noise C-DIR readout allows operation at lower MCP gain and higher point source count rates, and the lifetime advantages conferred by ALD technology extends the useful detector operation by over an order of magnitude.

MCP performance is limited by the photocathodes efficiency. In the NUV semitransparent solar-blind CsTe photocathodes can now routinely achieve >30% QE and GaN operated in reflection mode (possibly usable for spectroscopy) has been measured at 80% at 1200Å but only 20% thus far in semi-transparent mode necessary for imaging. CsI is still likely the current photocathode of choice for FUV vacuum photocathodes, capable of >40% QE in the 400-1200Å range. But developments in III-nitride materials are likely in the medium term.

A temporal resolution of 0.01 second for data imposed by science and/or spacecraft requirements is straightforwardly achievable by all of the candidates detector technologies.

**3.5 Gratings**
The grating technology is well developed and grating configurations are likely to be similar to those employed in recent missions. First order holographic gratings deliver excellent efficiency (~60% peak) and low scatter. However, the highest resolving powers available are a few tens of thousands and echelle systems are still required to deliver the highest spectral resolution required. The challenge is to produce systems with the low scatter achieved by 1$^{st}$ order gratings. New designs of low order echelle gratings with magnifying cross dispersers show promise.

**3.6 Additional elements**
Integral field spectroscopy: We could dedicate a small fraction of the large focal plane for an integral field spectrometer. By limiting the short wavelength to ~180 nm this could be achieved with a coherent bundle of fused silica fibers. To go to much shorter wavelengths it would be necessary of have fibers of $MgF_2$ or microshutter arrays, requiring technological development.

Spectropolarimetry: Magnesium fluoride is bi-refringent allowing the separation of an image into its two perpendicular polarizations. This would need to be inserted into two spectrographs, to produce spectra of these.

*In summary:*
We present a compelling and relevant set of science investigations leading to the probing of the building blocks of life in the Universe. These investigations will be possible and enabled if the European Ultraviolet-Visible Observatory (EUVO) moves forward to its next development phase.